\documentclass[twocolumn,aps,prd,showpacs]{revtex4}

\usepackage{graphics,graphicx}
\usepackage{amssymb,amsfonts}

\def\noi{\noindent}

\newcommand{\sect}[1]{Sec.\,#1}
\def\nq{\hspace*{-1em}}
\def\nqq{\hspace*{-2em}}

\def\nhh{\hspace*{-0.3em}}
\def\cm{\hspace*{1cm}}
\def\inch{\hspace*{1in}}

\def\ten#1{\mbox{$\cdot 10^{#1}$}}

\def\Jl#1#2{{\it #1\/} {\bf #2},\ }

\def\ApJ#1 {\Jl{Astroph. J.}{#1}}
\def\CQG#1 {\Jl{Class. Quantum Grav.}{#1}}
\def\DAN#1 {\Jl{Dokl. AN SSSR}{#1}}
\def\GC#1 {\Jl{Grav. \& Cosmol.}{#1}}
\def\GRG#1 {\Jl{Gen. Rel. Grav.}{#1}}
\def\JETF#1 {\Jl{Zh. Eksp. Teor. Fiz.}{#1}}
\def\JETP#1 {\Jl{Sov. Phys. JETP}{#1}}
\def\JHEP#1 {\Jl{JHEP}{#1}}
\def\JMP#1 {\Jl{J. Math. Phys.}{#1}}
\def\NPB#1 {\Jl{Nucl. Phys.}{B\ #1}}
\def\NP#1 {\Jl{Nucl. Phys.}{#1}}
\def\PLA#1 {\Jl{Phys. Lett.}{#1A}}
\def\PLB#1 {\Jl{Phys. Lett.}{#1B}}
\def\PRD#1 {\Jl{Phys. Rev.}{D\ #1}}
\def\PRL#1 {\Jl{Phys. Rev. Lett.}{#1}}

\def\lal{&& {}\nhh}
\def\eq{Eq.\,}
\def\eqs{Eqs.\,}
\def\beq{\begin{equation}}
\def\eeq{\end{equation}}
\def\bear{\begin{eqnarray}}
\def\bearr{\begin{eqnarray} \lal}
\def\ear{\end{eqnarray}}
\def\earn{\nonumber \end{eqnarray}}
\def\nn{\nonumber\\ {}}
\def\nnv{\nonumber\\[5pt] {}}
\def\nnn{\nonumber\\ \lal }
\def\nnnv{\nonumber\\[5pt] \lal }

\def\eql{&\! = &\!}

\def\dst{\displaystyle}
\def\tst{\textstyle}
\def\fracd#1#2{{\dst\frac{#1}{#2}}}
\def\fract#1#2{{\tst\frac{#1}{#2}}}
\def\Half{{\fracd{1}{2}}}
\def\half{{\fract{1}{2}}}

\def\e{{\,\rm e}}
\def\d{\partial}

\def\sign{\mathop{\rm sign}\nolimits}

\def\const{{\rm const}}
\def\eps{\varepsilon}

\def\DAL{\mathop{\raisebox{-.5pt}{$\Box$}}\nolimits}

\newcommand{\lims}[1]{\mathop{#1}\limits}
\newcommand{\limr}[4]{\,\raisebox{#1}{${\lims{#2}^{#3}_{#4}}$}\,}

\def\lsim{\ \limr{3pt}{<}{}{\tst\sim}\ }

\def\mn{_{\mu\nu}}
\def\MN{^{\mu\nu}}
\def\mN{_\mu^\nu}
\def\nM{_\nu^\mu}
\def\cK{{\cal K}}
\def\cV{{\cal V}}

\def\oR{{\overline R}}

\def\kappa{\varkappa}

\def\wt{\widetilde}
\def\tg{{\wt g}}
\def\tR{{\wt R}}

\def\H{{\mathbb H}}
\def\R{{\mathbb R}}

\def\Lambdaef{\Lambda_{\rm eff}}

\def\sss{\scriptscriptstyle}
\def\aE{a_{\sss\rm E}{}}
\def\mD{m_{\sss D}}

\def\dotaE{{\dot a}_{\sss\rm E}{}}

\begin{document}

\title{Self-stabilization of extra dimensions}

\author{K.A. Bronnikov}
\affiliation
    {Center for Gravitation and Fundamental Metrology, VNIIMS, 46 Ozyornaya St., Moscow, Russia;
     Institute of Gravitation and Cosmology, PFUR, 6 Miklukho-Maklaya St., Moscow 117198, Russia.
         E-mail: kb20@yandex.ru}

\author{S.G. Rubin}
\affiliation
    {Moscow State Engineering Physics Institute, 31 Kashirskoe Sh., Moscow 115409, Russia.
    E-mail: sergeirubin@list.ru}

\begin{abstract}
{We show that the problem of stabilization of extra dimensions in
 Kaluza-Klein type cosmology may be solved in a theory of gravity involving
 high-order curvature invariants. The method suggested (employing a
3 slow-change approximation) can work with rather a general form of the
 gravitational action. As examples, we consider pure gravity with
 Lagrangians quadratic and cubic in the scalar curvature and some more
 complex ones in a simple Kaluza-Klein framework. After a transition to
 the 4D Einstein conformal frame, this results in effective scalar field
 theories with certain effective potentials, which in many cases possess
 positive minima providing stable small-size extra dimensions. Estimates
 made in the original (Jordan) conformal frame show that the problem of a
 small value of the cosmological constant in the present Universe is
 softened in this framework but is not solved completely.}

\bigskip
Short title: Self-stabilization of extra dimensions


\end{abstract}
\maketitle


\section{Introduction}

   Modern cosmology is able to explain many properties of our Universe due
   to a substantial progress in experimental methods and technology and
   development of new theoretical models. Nevertheless, there is quite a
   number of long-standing and challenging unsolved problems. We are going
   to discuss some of them. The first one, known for two decades,
   concerns the origin and form of the inflaton potential \cite{Lindebook}
   able to provide a sufficient primordial inflation satisfying the
   observational constraints. Despite numerous attempts, see, e.g.,
   \cite{Hybrid, Massive}, the problem is yet to be solved. Another problem
   is related to the recently observed accelerated expansion of the Universe
   whose origin is usually ascribed to some unknown and invisible dark
   energy, which is likely to be identified with an extremely small but
   nonzero vacuum energy density, or cosmological constant. Certain problems
   of theoretical origin are common to all modern theories, employing the
   idea of extra dimensions, among which well-known are M-theory,
   supergravities, diverse Kaluza-Klein and brane-world scenarios. A
   requirement which inevitably accompanies such investigations is that the
   extra dimensions should be invisible since our space-time looks
   4-dimensional at least at scales over $10^{-17}$ cm. In models where the
   extra dimensions are assumed to be compact, this is a severe restriction
   on their size. Moreover, to account for the observed constancy or nearly
   constancy of the fundamental physical constants (the gravitational
   constant $G$, the fine structure constant $\alpha$ and others
   \cite{constants}), this size should be constant and stable, and therefore
   a mechanism of stabilization of the extra dimensions is required in any
   scenario that involves them.

   The most popular way of reaching such a stabilization is to invoke a
   scalar field of non-geometric origin with an appropriate potential. This
   new entity seems somewhat artificial because scalar fields already exist
   in this framework: the metric tensor components corresponding to the
   extra dimensions behave as scalar and vector fields in our 4D space. It
   is, however, difficult to obtain a suitable form of the potential for
   such fields \cite{Kolb, Mazumdar}. To improve the situation, additional
   non-gravitational fields are invoked thus returning us to the starting
   point.

   On the other hand, the gravitational action, in general, includes terms
   containing nonlinear functions of the Ricci scalar as well as other
   curvature invariants. Such nonlinear terms inevitably appear if one takes
   into account quantum phenomena \cite{gmm,bird,Don}. In fact, the first
   model of inflation was based on this idea \cite{Star80}. Gravity
   represents a wide choice of different forms of action since even
   renormalization of non-gravitational quantum fields against a curved
   background yields various curvature terms with indefinite coefficients,
   to say nothing of presently unknown classical ``traces'' of quantum
   gravity and Casimir-like contributions from a nontrivial space-time
   topology (if any). This wealth of possibilities has a negative imprint.
   Indeed, only one form of the gravitational action among many is realized
   in our Universe, and it is a challenging problem to choose the correct
   one.

   In this paper, we study curvature-nonlinear gravity in a space-time
   manifold with an arbitrary number of extra dimensions. Our aim is
   threefold. First, we intend to demonstrate that even the simplest version
   of nonlinear multidimensional gravity (with the Lagrangian $F(R)$, $R$
   being the Ricci scalar) leads to a scalar field with potential and
   kinetic terms having promising shapes. Second, we are going to point out
   some particular examples of minima of the effective potential $V$ with
   positive $V = V_{\min}$, which are, in principle, able to describe
   the present accelerated expansion of the Universe (certainly in a very
   rough approximation). Third, we try to find out whether or not, or to
   which extent, the well-known cosmological constant problem can be
   solved in the present class of theories.

   In our study we use the slow-change approximation, assuming smallness
   of all derivatives, in a sense discussed below, and smallness of energy
   densities compared to the Planck scale. Such assumptions are rather
   widespread in studies of inflationary scenarios. This method allows
   dealing with rather general forms of the gravitational action and
   extended parameter ranges.  With its help, new minima of the potential
   which can provide stability and smallness of the extra dimensions are
   found.

   The paper is organized as follows. In \sect \ref{S2} we consider $F(R)$
   theories of gravity in multidimensional space-times where the extra
   dimensions comprise a constant-curvature space with a scale factor
   $b(x)$, depending on the external (observed) coordinates. Integrating out
   the extra dimensions in the action and applying the slow-change
   approximation, we obtain an effective scalar-tensor theory in 4
   dimensions with a single scalar field $b(x)$ and specific forms of its
   potential and kinetic terms. It should be noted that the complexity of
   the kinetic term can also affect the system dynamics and qualitatively
   change its properties \cite{Kessence, Ru04}.

   We also show that our approximation agrees with the conventional approach
   that reduces $F(R)$ gravity to $D$-dimensional Einstein gravity coupled
   to a scalar field with a potential related to $F(R)$.

   \sect \ref{S2a} is devoted to numerical estimates concerning possible
   stationary states of the extra dimensions. We briefly discuss the role of
   conformal frames, distinguishing the fundamental frame in which the
   theory is originally formulated (saying what happens as a matter of
   fact) and the frame used to interpret the observations (showing what we
   can see). The choice of the latter depends on the properties of references
   employed in our measurements \cite{pictures}. We here restrict ourselves
   to the simplest and maybe the most natural assumption, that the
   observational frame coincides with the fundamental (Jordan) one. Under
   this assumption, it turns out that the smallness of the present
   cosmological constant remains a problem though less severe than in
   conventional cosmology.

   In \sect \ref{S3} we use our method for seeking stable states of the
   extra dimensions in quadratic multidimensional gravity, with $F(R) = R +
   cR^2 -2\Lambda$ where $c$ and $\Lambda$ are constants. This theory has
   been previously discussed in Ref.\,\cite{Zhuk}, where local minima of the
   potential $V$ with $V_{\min} < 0$ were found at values of $R$ corresponding
   to $dF/dR > 0$. In the cosmological context, such minima can only lead to
   models with the anti-de Sitter space AdS$_4$. We confirm this result but
   also find new minima of the potential in the unusual range where $dF/dR <
   0$; for some of them, $V_{\min} > 0$, which leads to de Sitter cosmology.

   \sect \ref{S4} discusses other choices of the initial Lagrangian.
   Thus, with $F(R)$ taken in the form of a cubic polynomial, we find minima
   of the effective potential with $V_{\min} >0$ in the region where
   $dV/dR > 0$. Then we demonstrate that our method can be successfully
   applied to a much wider class of multidimensional theories of gravity,
   e.g., those containing the Ricci tensor squared and the Kretschmann
   scalar. \sect\ref{S5} contains some concluding remarks.

\section{$F(R)$ theory in $D$ dimensions \label{S2}}

\subsection{Basic equations}

    We consider a $(D = d_0 + d_1)$-dimensional manifold with the metric
\beq                                            \label{ds}
    ds^2 = g\mn dx^\mu dx^\nu + \e^{2\beta(x)} b_{ab} dx^a dx^b
\eeq
    where the extra-dimensional metric components $b_{ab}$ are independent of
    $x^{\mu}$, the observable space-time coordinates. The relevant case
    is certainly $d_0 = 4$ but, in \eqs (\ref{Riem})--(\ref{act1}), we keep
    $d_0$ arbitrary for generality.

    The $D$-dimensional Riemann tensor has the nonzero components
\bear
    R\MN{}_{\rho\sigma} \eql \oR\MN{}_{\rho\sigma},       \label{Riem}
\nn
    R^{\mu a}{}_{\nu a} \eql \delta^a_b\, B\nM, \cm
    B\nM := \e^{-\beta} \nabla_\nu (\e^\beta \beta^{\mu}),        
\nn
    R^{ab}{}_{cd} \eql
      \e^{-2\beta} \oR^{ab}{}_{cd} + \delta^{ab}{}_{cd} \beta_\mu\beta^\mu,
\ear
    where capital Latin indices cover all $D$ coordinates, the bar marks
    quantities obtained from $g\mn$ and $b_{ab}$ taken separately,
    $\beta_{\mu}\equiv\d_{\mu}\beta$ and $\delta^{ab}{}_{cd}\equiv
    \delta_{c}^{a}\delta_{d}^{b}-\delta_{d}^{a}\delta_{c}^{b}$. The
    nonzero components of the Ricci tensor and the scalar curvature are
\bear
    R\mN \eql \oR\mN + d_1\, B\mN,          \label{Ric}           
\nn
    R_a^b \eql \e^{-2\beta} \oR_a^b
                    + \delta_a^b [ \DAL \beta + d_1 (\d{\beta})^2 ],
\nnv
    R \eql \oR [g] + \e^{-2\beta }
               \oR [b] + 2d_1 \DAL \beta + d_1(d_1+1) (\d{\beta})^2,
\nnn
\ear
    where $(\d{\beta})^2 \equiv \beta_{\mu}\beta^{\mu}$,
    $\DAL = g\MN\nabla_\mu \nabla_\nu$ is the $d_0$-dimensional d'Alembert
    operator while $\oR[g]$ and $\oR [b]$ are the Ricci scalars
    corresponding to $g\mn$ and $b_{ab}$, respectively.

    Suppose now that $b_{ab}$ describes a $d_1$-dimensional space of
    nonzero constant curvature, i.e., a sphere ($k=1$) or a compact
    $d_1$-dimensional hyperbolic space \cite{Lob} ($k = -1$) with a
    fixed curvature radius $r_0$ normalized to the $D$-dimensional
    analogue $\mD$ of the Planck mass, i.e., $r_0 = 1/\mD$ (we use the
    natural units, with the speed of light $c$ and Planck's constant $\hbar$
    equal to unity). We have
\bear
    \oR^{ab}{}_{cd} \eql k\, \mD^2\,\delta^{ab}{}_{cd},         \label{r0}
            \cm   \oR_a^b = k\, \mD^2\, (d_1-1) \delta_a^b,
\nn
    \oR [b] \eql k\, \mD^2\, d_1 (d_1-1) = R_b.
\ear
    The scale factor $b(x) \equiv \e^{\beta}$ in (\ref{ds}) is thus kept
    dimensionless; $R_b$ has the meaning of a characteristic curvature scale
    of the extra dimensions.

    Consider, in the above geometry, a sufficiently general
    curvature-nonlinear theory of gravity with the action
\beq                                                         \label{act1}
     S = \Half \mD^{D-2} \int\sqrt{^{D}g}\,d^{D}x\,[F(R) + L_m],
\eeq
    where $F(R)$ is an arbitrary smooth function, $L_m$ is a matter
    Lagrangian and ${^D}g = |\det(g_{MN})|$. The extra coordinates are
    easily integrated out, and the action is reduced to $d_0 = 4$ dimensions:
\beq
     S = \Half \cV [d_1]\,\mD^2 \int\sqrt{^4g}\,d^{4}x\,
            \e^{d_1\beta}\,[F(R)+L_m],                      \label{act2}
\eeq
    where $^{4}g = |\det(g\mn)|$ and $\cV[d_1 ]$ is the volume of a
    compact $d_1$-dimensional space of unit curvature.

\subsection {Slow-change approximation. The Einstein frame}

    \eq (\ref{act2}) describes a 4D theory which is nonlinear in curvature
    and, moreover, contains a non-minimal coupling between the effective
    scalar field $\beta$ and the curvature. Let us simplify it in the
    following manner:

\medskip\noi
{\bf (a)} Express everything in terms of 4-dimensional
    variables and $\beta(x)$; note that now
\bearr                                                      \label{R4}
        R = R_4 + \phi + f_1,
\nnnv
        R_4 = \oR [g],      \qquad
            f_1 = 2d_1 \DAL \beta + d_1(d_1+1)(\d{\beta})^2,
\ear
    and we have introduced the effective scalar field
\beq                                         \label{phi}
        \phi (x) = R_b \e^{-2\beta (x)}
        = kd_1(d_1-1)\mD^2\, \e^{-2\beta (x)}.
\eeq
    Recall that we have $k = \pm 1$ for positive and negative curvature in
    $d_1$ extra dimensions, respectively, so that $\phi$ has different signs
    in these cases by definition.

\medskip\noi
{\bf (b)} Suppose that all quantities are slowly varying, i.e., consider
    each derivative $\d_{\mu}$ (including those in the definition of $\oR$)
    as an expression containing a small parameter $\eps$; neglect all
    quantities of orders higher than $O(\eps^2 )$ (see \cite{Don}).

\medskip\noi
{\bf (c)} Perform a conformal mapping leading to the Einstein conformal
    frame, where the 4-curvature appears to be minimally coupled to the
    scalar $\phi$.

\medskip
    In the decomposition (\ref{R4}), both terms $f_1$ and $R_4$ are regarded
    small in our approach, which actually means that all quantities,
    including the 4D curvature, are small compared to the $D$-dimensional
    Planck scale. So the only term which is not small is $\phi$, and we can
    use a Taylor decomposition of the function $F(R) = F(\phi + R_4 + f_1)$:
\beq                                                       \label{Fapprox}
    F(R) = F(\phi + R_4 + f_1 )\simeq F(\phi) + F'(\phi)\cdot(R_4 +f_1 )+...,
\eeq
    with $F'(\phi)\equiv dF/d\phi$.
    Substituting it into \eq (\ref{act2}), we obtain up to $O(\eps^2)$
\bearr
     S = \Half {\cV[d_1 ]}\, \mD^2                           \label{act3}
          \int \sqrt{^{4}g}\,d^{4}x\, \e^{d_1 \beta}
             [F'(\phi)R_4       \cm
\nnn \inch
         + F(\phi) + F'(\phi) f_1 + L_m],
\ear
    where $\beta$ is related to $\phi$ according to (\ref{phi}). The
    expression (\ref{act3}) is typical of a scalar-tensor theory
    (STT) of gravity in a Jordan frame.

    To find stationary points, it is helpful to pass on to the Einstein
    frame. After the conformal mapping
\beq                                                       \label{trans-g}
    g\mn \ \mapsto \tg\mn = |f(\phi)| g\mn, \cm
            f(\phi) =  \e^{d_1\beta}F'(\phi),
\eeq
    with the corresponding transformation of the scalar curvature
\beq                                        \label{trans-R}
    R_4 = |f| \biggl(\tR_4 + \frac{3}{f} {\wt\DAL} f
                    - \frac{3}{2f^2} (\wt\d f)^2 \biggr)
\eeq
    (the tilde marks quantities obtained from or with $\tg\mn$),
    the action (\ref{act3}) acquires the form
\bearr                                                       \label{Sgen}
    S = \Half {\cV [d_1]}\, \mD^2  \int \sqrt{\tg}\, d^4 x
       \biggl\{ [\sign F'(\phi)]\ [\tR_4 + K]
\nnn \cm \cm
       - V(\phi) + \frac{\e^{-d_1\beta}}{F'(\phi)^2}L_m \biggr\}
\ear
    with the kinetic ($K$) and potential ($V$) terms
\bear
        K \eql \frac{3}{2}\frac{(\d f)^2 }{f^2 }-2d_1
                \frac{f^{,\mu} \beta_\mu}{f} + (\d\beta)^2,
\nn
        V(\phi) \eql -\e^{-d_1 \beta} \, F(\phi)/F'(\phi)^2 .
\ear

    It remains to express everything in terms of a single scalar variable,
    say, $\phi$. We can write the action (\ref{Sgen}) in the form
\bear
     S \eql \frac{\cV[d_1]}{2} \mD^2 \int \sqrt{\tg}\, (\sign F') L,
\nn
     L \eql \tR_4 + \Half K_{\rm Ein}(\phi) (\d\phi)^2
                        - V_{\rm Ein}(\phi) + {\wt L}_m,   \label{Lgen}
\nnn \ \
          {\wt L}_m = (\sign F')\frac{\e^{-d_1\beta}}{F'(\phi)^2} L_m;
\\
     K_{\rm Ein}(\phi) \eql                                \label{Kgen}
        \frac{1}{2\phi^2} \left[
            6\phi^2 \biggl(\frac{F''}{F'}\biggr)^2\!
            -2 d_1 \phi \frac{F''}{F'} + \Half d_1 (d_1{+}2)\right],
\nnn
\\
     V_{\rm Ein}(\phi) \eql - (\sign F')
        \left[\frac{|\phi|\mD^{-2}}{d_1 (d_1 -1)}\right]^{d_1/2}
                \frac{F(\phi)}{F'(\phi)^2 }            \label{Vgen}
\ear
    In (\ref{trans-R})--(\ref{Lgen}), the indices are raised and lowered
    with $\tg\mn$; everywhere $F = F(\phi)$ and $F' = dF/d\phi$.

    \eqs (\ref{Lgen})--(\ref{Vgen}) are valid for both positive ($\phi > 0$)
    and negative ($\phi < 0$) curvature of the extra dimensions. Minima of
    the potential $V_{\rm Ein}$, defined in the Einstein frame, determine
    stationary points of the scalar field, and they remain to be stationary
    points after a transition to any other conformal frame, including the
    original Jordan frame. It is the Einstein frame that determines the
    scalar field behavior since its dynamics near an extremum is governed
    directly by the potential $V(\phi)$ and only implicitly by the metric.

    If $\phi$ resides at a minimum of the potential $V_{\rm Ein}$, this
    potential turns into an effective cosmological constant. Being applied
    to the present cosmological epoch, it can determine the observable dark
    energy density that drives the accelerated expansion of the Universe.
    Alternatively, in principle, it may drive inflation in the early Universe.

    We have achieved one of our goals: the potential (\ref{Vgen}) valid for
    any function $F(R)$ looks quite complex to have some nontrivial extrema.
    In addition, we have obtained rather a complex form of the kinetic term
    (\ref{Kgen}). Its properties are known to be as important for the field
    dynamics as the shape of the potential. For example, as shown in
    Ref.\,\cite{Ru04}, zeros and singular points of the kinetic term may be
    responsible for stable states of a scalar field.

    In our expressions, both the effective potential (\ref{Vgen}) and
    the kinetic term (\ref{Kgen}) are singular at the values of $\phi$
    where $f(\phi)\sim F'(R)$ [the factor before $R_4$ in the Jordan-frame
    action (\ref{act3})] is zero. Unlike many papers restricted to $F'(R) >
    0$, we also include models with $F'(R) < 0$. As will be seen below, this
    opens new promising possibilities such as new minima of the effective
    potential at which the extra dimensions may be stabilized.  Moreover,
    models with conformal continuations \cite{vac4,hog1}, which unify
    regions with $F' > 0$ and $F' < 0$, are possible; these regions
    correspond to different Einstein-frame manifolds but turn out to be
    smoothly connected in the Jordan frame. Though, such models are likely
    to be unstable, as follows from the experience of dealing with different
    solutions of scalar-tensor theories \cite{instab}. Our main interest
    here is in other values of $\phi$, namely, those at which it may be
    stabilized.

    In (\ref{Lgen})--(\ref{Vgen}) we have actually changed the sign of the
    Lagrangian in the case $F' < 0$; to preserve the attractive nature of
    gravity for ordinary matter, the matter Lagrangian density should appear
    with an unusual sign from the beginning. As a result, the sign of the
    whole action of gravity and matter will be unusual, without any effect
    on the matter equations of motion, and the conventional form of the
    effective Einstein equations at a stationary value of $\phi$ will also
    be preserved. Only the action of the $\phi$ field itself can be unusual,
    according to \eqs (\ref{Kgen}), (\ref{Vgen}). It should be noted that
    the common sign of the total action does not affect quantum transitions
    as well. Indeed, the transition amplitude is expressed in the path
    integral technique as $\int \exp{(iS[q])}Dq$ where $q(t)$ is some
    dynamical variable. The transition probability $\int \int
    \exp{(iS[q_{1}] - iS[q_{2}])}Dq_{1}Dq_{2}$ is invariant under the
    substitution $i\to -i$ (with interchanging the integration variables
    $q_1$ and $q_2$).

\subsection {Comparison with the conventional approach}

    There is a well-known conformal mapping \cite{f(R)} bringing the theory
    (\ref{act1}) to the form of general relativity with a minimally coupled
    scalar field $\chi$ that replaces the additional degrees of freedom
    connected with $R$:
\bear   \nq                        \label{trans-h}
     g_{AB} \eql \e^{-2\omega(\chi)} h_{AB},
\nn
     \chi \eql \pm\sqrt{\frac{D-1}{D-2}} \ln |F_R|,
\quad\
     \e^\omega(\chi) = |F_R|^{1/(D-2)},
\ear
    with $F_R \equiv dF/dR$. This results in a $D$-dimensional Einstein
    frame, with the action
\bearr  \nq                                                    \label{act-h}
     S = S_{DE}[h] = \Half \mD^{D-2} \int\sqrt{^{D}h}\,d^{D}x
\nnn \times
     \biggl\{(\sign F_R)\bigl[R[h] + (\d\chi)^2_h]
        - 2 V(\chi) + L_m \e^{-D\omega}\biggr\},
\nnn\cm
        2 V(\chi) = \e^{-D\omega} (RF_R - F),
\ear
    where $R[h]$ is the Ricci scalar obtained from $h_{AB}$,
    $^D h = |\det (h_{AB})|$ and $(\d\chi)^2_h = h^{AB}\chi_{,A}\chi_{,B}$;
    the initial Ricci scalar $R$ is considered as a function of $\chi$.

    (An even more general transformation is known \cite{maeda}, bringing a
    $D$-dimensional theory of gravity whose Lagrangian contains an arbitrary
    function of two variables, $f(R,\phi)$, to an Einstein frame with two
    minimally coupled scalar fields, $\phi$ and the one similar to $\chi$.
    Maeda's method \cite{maeda} generalizes the corresponding mappings known
    for both scalar-tensor \cite{wagon} and $f(R)$ \cite{f(R)} theories of
    gravity.)

    The theory (\ref{act-h}) may be further reduced to 4 dimensions and
    brought with one more conformal mapping, now depending on $\beta(x)$
    and similar to (\ref{trans-g}), to a 4-dimensional Einstein frame.
    Reduction in terms of the metric $h_{AB}$ gives
\bearr                                                      \label{S_h}
     S = S_{4J}[h]
       = \Half \mD^{D-2} \cV[d_1]\int d^4 x \,\sqrt{^4 h}\,\e^{d_1 \alpha}
\nnn \times
         \biggl\{(\sign F_R) \biggl[R_4[h] + (\d\chi)^2_h
              - d_1(d_1-1) (\d\alpha)^2_h
\nnn \cm\cm                                                    
          - \e^{-2 \alpha} R_b \biggr]
             - 2 V(\chi) + L_m \e^{-D\omega} \biggr\},
\nnn
      \alpha := \beta + \omega,
\ear
    where $^4h = |\det (h\mn)|$. After a further transformation,
\beq
      h\mn = \e^{-d_1\alpha} \gamma\mn,                        
\eeq
    we arrive at a theory defined in the 4D Einstein frame with the metric
    $\gamma\mn$. The action is
\bearr  \nq                                                 \label{S_gamma}
     S = S_{4E}[\gamma]= \Half \mD^{D-2} \cV[d_1]              
         \int\sqrt{\gamma}\,d^4 x\,
\nnn \nq \times
     \biggl\{(\sign F_R)\biggl[R[\gamma] + (\d\chi)^2_\gamma
          + \half d_1(d_1+2) (\d\alpha)^2_\gamma
\nnn  \nqq
          - \e^{-(d_1+2)\alpha} R_b \biggr]
        - 2 V(\chi)\e^{-d_1\alpha} + L_m \e^{-D\omega-d_1\alpha}\biggr\},
\ear
    with $\gamma = |\det(g\mn)|$ and
    $(\d\chi)^2_\gamma = \gamma\MN \chi_{,\mu}\chi_{,\nu}$.

    Evidently, under the conditions of our slow-change approximation, the
    action (\ref{S_gamma}) should lead to the same results as (\ref{Sgen}).
    Let us confirm that it is indeed the case, assuming in (\ref{S_gamma})
    that all derivatives $\d_\mu$ involve a small parameter. We have, as
    before, (\ref{Fapprox}) and similarly
\beq                                                    \label{F'_approx}
       F_R = F'(\phi) + F''(\phi) (R_4 + F_1) + ...,
\eeq
    where $R_4 = R_4[g]$. Then, comparing the transitions $g_{AB} \to g\mn
    \to \tg\mn$ in (\ref{trans-g}) and $g_{AB} \to h_{AB} \to h\mn \to
    \gamma\mn$, we see that the two Einstein-frame metrics $\tg$ and
    $\gamma$ are related by
\beq                                                        \label{g_compar}
     \tg\mn = \frac{F'(\phi)}{F_R} \gamma\mn =
     \biggl[ 1 + \frac{F''(\phi)}{F'(\phi)}(R_4 + f_1)+...\biggr]\gamma\mn
\eeq
    and coincide in the leading order of magnitude. The leading terms of
    the actions (\ref{Lgen}) and (\ref{S_gamma}), represented by their
    potential and matter terms, should then also coincide, and it is easy to
    confirm that it is really the case (the term $RF_R$ in (\ref{S_gamma})
    transforms to $\phi F'(\phi)$ and is cancelled by the term with $\R_b$,
    leading to the expression (\ref{Vgen}) for the $\phi$ field potential).

    It is also seen that both fields $\chi$ and $\alpha$ are reduced, in the
    same approximation, to the $\phi$ field defined in \eq (\ref{phi}):
\beq
       \alpha_{,\mu} \approx \beta_{,\mu}
            + \frac{1}{D-2}\frac{F''}{F'} \phi_{,\mu},\quad
       \chi_{,\mu} \approx \sqrt{\frac{D-1}{D-2}} \frac{F''}{F}\phi_{,\mu},
\eeq
    and $\beta_{,\mu} = -\phi_{,\mu}/(2\phi)$. As a result, the whole
    kinetic term takes precisely the form (\ref{Kgen}). So the action
    (\ref{Lgen}) is completely restored.

    One can note, however, that our approximation, being applied from the
    very beginning, not only makes things simpler but is more universal: it
    can deal with actions more general than (\ref{act1}) (see
    \sect\ref{S4}).

\section {Effective constants: some estimates \label{S2a}}

\subsection {Conformal frames and requirements to the model}

    To relate the theory to Nature, it is necessary to specify which
    conformal frame is to be confronted to observations.

    In our problem setting, the Jordan frame described above appears to be
    fundamental since the initial theory is formulated in it, and the quantum
    effects that give rise to the nonlinearity of gravity are also supposed
    to take place there.

    It is, however, quite unnecessary to believe that this Jordan frame
    corresponds to the observed picture, in which the properties of our
    measurement instruments and the standards of the relevant physical units
    do not change in space and time \cite{pictures}. The latter is thus the
    frame in which the basic atomic units (depending, above all, on fermion
    masses) are constant, if certainly they are all constant simultaneously.
    How sensitive is the matter Lagrangian to conformal mappings is even
    well seen from the coefficient before $L_m$ in \eq (\ref{Lgen}), to say
    nothing of a nontrivial metric dependence of matter fields entering into
    $L_m$.

    Since we here do not specify how fermions are included into the theory,
    it is only possible to assume which is the observational frame.
    We will here dwell upon, probably, the most natural choice and make
    some estimates in the Jordan frame thus assuming that the basic and
    observational frames coincide.

    As regards the Einstein frame in which the action takes the form
    (\ref{Sgen}), we simply use it as a technical trick to determine
    stationary states.

    In any observational frame, in the expected stationary state, the
    gravitational action has the approximate Einstein-Hilbert form
\beq                                                            \label{S_GR}
       S_{\rm GR}= \Half m_4^2 \int \sqrt{^4g}\,d^4 x\,(R_4^* -2\Lambdaef),
\eeq
    where $m_4 = (8\pi G_N)^{-1/2}$ is the 4D Planck mass, $G_N$ is the
    effective Newtonian constant and $R^*_4$ is the observable 4D curvature
    (for our choice, $R_4^* = R_4$).
    To be consistent with our approach and with observational constraints,
    this stationary state should satisfy the following requirements:
\begin{description}
\item[(i)]
    A classical space-time description should be admissible, i.e., the
    true size of the extra dimensions should exceed the true
    $D$-dimensional Planck length $1/\mD$, the fundamental length scale
    of the theory:
\beq                                                          \label{clas}
    \mD b_0 = \e^{\beta _0} \gg 1, \quad
                          {\rm or} \quad  \phi/\mD^2  \ll d_1(d_1-1).
\eeq
\item[(ii)]
    The slow-change approximation should work, which, for
    $\phi = \const$, reduces to the requirement $R_4 \ll m_D^2$.
\item[(iii)]
    The observed size $b^*$ of the extra dimensions can be much larger
    than the Planck length $l_4 = 1/m_4 \approx 8\ten{-33}$ cm, but not
    larger than about $10^{-17}$ cm, which corresponds to the TeV energy
    scale.
\item[(iv)]
    The predicted effective cosmological constant $\Lambdaef$
    should be very small to conform to the observations:
\beq                                                       \label{Lam}
        \Lambdaef/m_4^2 \sim 10^{-120}
\eeq
    (the exponent is $-120$ instead of more conventional $-122$ because
    the definition of $m_4$ involves $8\pi$).
\end{description}

    Obtaining such a small value as (\ref{Lam}) is one of the well-known
    problems of theoretical cosmology since it is hard to explain without
    fine tuning why $\Lambda m_4^2$, generally associated with vacuum energy
    density, is so many orders of magnitude smaller than the characteristic
    energy densities inherent to the known physical interactions (e.g., the
    Planck density $m_4^4$ for gravity).

\subsection{Estimates in the Jordan frame}

    Thus we assume that observations are performed in the Jordan frame.
    The corresponding action (\ref{act3}) is approximated by (\ref{S_GR})
    with the effective constants
\beq                                            \label{G_NJ}
    \frac{1}{8\pi G_N} = m_4^2 =
                    \cV[d_1]\, \mD^{D-2} \, b_0^{d_1} F'_0,
    \quad\
        \Lambdaef = - \frac{F(\phi_0)}{2 F'_0},
\eeq
    where $b_0 = \e^{\beta (\phi_0)}/\mD = b_*$ is the observable size of
    the extra dimensions and $F'_0 = F'(\phi_0)$.

    \eq (\ref{G_NJ}) leads to the following relation between the
    dimensionless quantities $b_0 \mD$ and $b_0 m_4$:
\beq                                                        \label{mD,4}
        b_0^2 m_4^2 = \cV[d_1]\, F'_0\, (b_0 \mD)^{d_1+2}.
\eeq
    The factor $\cV[d_1]$ is of order unity; the same may be expected from
    the dimensionless quantity $F'_0$. By item (iii) above, $b_0^2 m_4^2
    \lsim 10^{30}$, and \eq (\ref{mD,4}) gives
\bearr                                                      \label{m4}
    b_0\mD \lesssim 10^{30/(d_1+2)},
\nnn
    m_4/\mD \sim (b_0 \mD)^{d_1} \lesssim 10^{15 d_1/(d_1+2)}.
\ear
    Thus $\mD$ not too much differs from $m_4$. It means, in particular,
    that our slow-change approximation (item (ii) above) works manifestly
    well in almost all thinkable circumstances since $R_4$ is the observable
    curvature.

    If $m_4/\mD$ is sufficiently close to unity, this approximation is
    even valid for the curvature characteristic of primordial inflation at
    the Grand Unification scale: thus, an a priori estimate is $R_4/m_4^2
    \sim 10^{-6}$. In particular models this inequality is strengthened.

    Thus, at chaotic inflation governed by the inflaton $\varphi$, the
    Ricci scalar is expressed in terms of the Hubble parameter $H$ and the
    potential $V (\varphi)$ as
\beq
     	R_{4} = 12 H^2;
\quad
	H \equiv \frac{\dot{a}}{a} \simeq \sqrt{\frac{8\pi G}{3}V(\varphi)},
			\label{eqH}
\eeq
    where the term with $\dot\varphi$ is omitted since $\varphi$ is assumed
    to be slowly rolling down the slope of the potential. Let us estimate
    $R_4$ in the quadrativ model of chaotic inflation, with
    $V = \half m^2\varphi^2$. The observational data indicate that
\beq
		H  \approx 10^{-6}m_{4}
\eeq
    during inflation, and so
\beq
		R_4/m_4^2 = 12 H^2/m_4^2 \sim 10^{-11}.
\eeq

    Due to the slow rolling, one may not worry about the smallness of $f_1$
    at the inflationary stage. The kinetic terms are, however, the largest
    at the end of inflation, at reheating, when $\varphi \sim  m_4$,
    while the slowly changing Hubble parameter becomes comparable with
    the inflaton mass, $H\sim m$, i.e., $m/m_4 \sim 10^{-6}$.
    Reheating is a result of quick oscillations of the inflaton around
    the minimum of $V$, where $\dot{\varphi}\sim m\varphi$, hence
    here we also obtain a small parameter:
\beq
		\dot{\varphi}/m_4^2 \sim 10^{-6},
\eeq
    which justifies our approximation provided $m_4/\mD < 10^6$. Otherwise
    our approximation starts to hold at later stages of the evolution, and
    it is manifestly good for the present-day Universe.

    Let us now return to the inequalities (\ref{m4}) and estimate
    $\Lambdaef$. The ratio $\Lambdaef/m_4^2$ may be expressed as follows:
\bearr                                                        \label{Lam1}
        - \frac{\Lambdaef}{m_4^2}
            = \frac{F(\phi_0)}{2\,\mD^2 F'_0}
                    \, \frac{\mD^2 b_0^2} {m_4^2 b_0^2}
\nnn \qquad
        = \frac{1}{2 F'_0}\, \frac{F(\phi_0)}{\mD^2}
               (F'_0 \cV[d_1])^{\fract{-2}{d_1+2}}
                (m_4 b_0)^{\fract{-2d_1}{d_1+2}}.
\nnn
\ear
    If we try to make this ratio small, the last factor in (\ref{Lam1}) can
    give at best about $10^{-30}$ (for sufficiently large $d_1$).
    The remaining 90 orders of magnitude must be gained due to unnatural
    smallness of the dimensionless quantity $F(\phi_0)/\mD^2$
    (actually, of the initial cosmological constant) and/or greatness of
    $F'_0$. The latter variant is, however, unsuitable since it will
    make impossible $b_0 \mD \gg 1$, see (\ref{mD,4}).

    As a result, the problem of fine tuning remains topical, though
    it should be noted that the very small value of $10^{-30}$ appears in
    this approach without any artificial effort.

\section{Quadratic gravity with a cosmological constant \label{S3}}

    In what follows we consider pure gravity ($L_m = 0$) and use the units
    $\mD =1$, thus dealing with dimensionless quantities.

    As the first and simplest example of multidimensional nonlinear gravity,
    consider (\ref{act1}) with the function
\beq                                                          \label{F2}
      F(\phi) = \phi + c\phi^2 - 2\Lambda, \cm c,\ \Lambda = \const.
\eeq
    Then \eqs (\ref{Kgen}) and (\ref{Vgen}) give the effective potential
\bearr
    V^{(2)}_{\rm Ein}(\phi) = - \sign(1+2c\phi)
        [d_1 (d_1 -1)]^{-d_1 /2}
\nnn \cm
    \times      |\phi|^{d_1/2}
         \frac{c\phi^2 + \phi - 2\Lambda} {(1+2c\phi)^2 }       \label{V}
\ear
    and the coefficient $K_{\rm Ein}$ of the kinetic term
\bearr
       K^{(2)}_{\rm Ein}(\phi)=
                \frac{1}{(1 + 2c\phi)^2 \phi^2 }
       \biggl[ c^2 \phi^2 (d_1^2 - 2d_1 + 12)
\nnn \cm \cm
       + d_1^2 c\phi
                + \frac{1}{4} d_1 (d_1 +2) \biggr].             \label{K}
\ear

\begin{figure}\label{FigZhuka}
    \includegraphics[angle=0,width=0.4\textwidth]{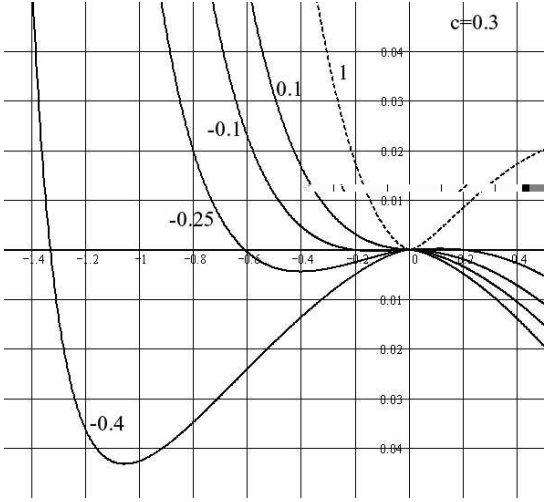}
\caption{The potential $V_{\rm Ein}(\phi)$ for $d_1=3$, $c=0.3$ in the range
    $\phi > -5/3$, where $F'(R) >0$, for a few values of $\Lambda$ (indicated
    near the curves). The dashed line ($\Lambda =1$) shows a minimum at $\phi
    = 0$. The values of $\varphi$ are expressed in the units $\mD =1$.}
\end{figure}

    In Ref.\,\cite{Zhuk}, the authors studied the case $F'(R) > 0 $ and
    showed analytically that a minimum of the potential stabilizing the
    extra dimensions is only possible with a negative energy density and
    a negative curvature $k = -1$. We confirm this result in our approach,
    see examples in Fig.\,1. The energy density corresponding to the minimum
    value of the potential is negative, which, in the cosmological context,
    corresponds to anti-de Sitter space-time.

    There is one more minimum of the potential $V_{\rm Ein}$ with $F' > 0$,
    existing in the range $\Lambda >0$ and located at the point $\phi =0$.
    The asymptotic $\phi \to 0$ corresponds to growing rather than
    stabilized extra dimensions: $b = \e^{\beta}\sim 1/\sqrt{|\phi|} \to
    \infty$. A model with such an asymptotic growth at late times may still
    be of interest if the growth is sufficiently slow and the size $b$ does
    not reach detectable values by now.

    Let us check whether it is possible to describe the modern state of
    the Universe by an asymptotic form of the solution for $\phi\to 0$
    as a spatially flat cosmology with the 4D Einstein-frame metric
\beq
        d{\wt s}{}^2_4 = dt^2 - \aE^2 (t) d\vec x{}^2,
\eeq
    where $\aE$ is the Einstein-frame scale factor. Two independent
    components of the Einstein-scalar equations for $\beta(t)$ and $\aE(t)$
    are
\bear                                                        
       \ddot \beta + 3\frac{\dotaE}{\aE} \dot{\beta}
                \eql \frac{2\Lambda} {d_1+1} \e^{-d_1\beta},
\nn
       3\frac{\dotaE^2}{\aE^2}                                \label{eq-a4}
           	\eql \frac{1}{4}d_1(d_1+1)\dot\beta{}^2
	    			+ \Lambda \e^{-d_1\beta}.
\ear
    We seek their solution at large $t$ in the form
\beq
       \aE(t) \propto t^p, \quad\   \e^{\beta} \approx (t/t_*)^q, \quad\
            p,\ q,\ t_* = \const >0,
\eeq
    and find that such a solution does exist with
\bearr                                                         \label{p,q}
        p = 3h, \cm q = 2/d_1,
\nnn
        h:= \sqrt{(d_1+1)/d_1},\cm   \Lambda t_*^2 = h^2 (9h-1).
\ear
    Passing over to the Jordan frame with
\beq
      ds^2_4 = d\tau^2 - a^2(\tau) d\vec x{}^2 = \frac{1}{f}\,d{\wt s}{}^2_4,
\eeq
    due to $F'(0) \approx 1$, we can put simply
    $f = \e^{d_1\beta} \approx (t/t_*)^2$. We obtain
\beq
        t/t_* \propto \e^{\tau/t_*},\quad\
      a(\tau) \propto \e^{(3h-1)\tau/t_*}, \quad\
      b(\tau) \propto \e^{2\tau/(d_1 t_*)}.
\eeq

    The external scale factor $a(\tau)$ grows exponentially, which conforms
    to modern observations if one properly chooses the constants. The
    internal scale factor $b(\tau)$ grows much slower for sufficiently large
    $d_1$, but the volume factor $\sim b^{d_1}$ grows approximately at the
    same rate as $a(\tau)$, which means that, e.g., the effective
    gravitational constant will change too rapidly, at nearly a Hubble rate,
    contrary to observations. We conclude that the model with $\phi\to 0$
    at late times is not very promising.

    We have also tried to find minima of the potential in another region,
    where $F'<0$, in a wide range of the parameters $c$ and $\Lambda$.

    Some examples of the behavior of the potential for the specific value
    $c=1.5$ are shown in Fig.\,2. Potentials with $\Lambda < 0$ show
    minima with $V_{\min} > 0$, so that $\Lambdaef > 0$, which should lead
    to cosmological models with de Sitter external space and stable extra
    dimensions. Such models again correspond to $\phi <0$, i.e., hyperbolic
    extra dimensions. By fine tuning of the initial constants $c$ and
    $\Lambda$ we can obtain the present-day values of the cosmological
    parameters.

    As in Fig.\,1, we also find a minimum at $\phi = 0$, whose properties
    have already been discussed.

\begin{figure}\label{Multyc1_5a}
    \includegraphics[angle=0,width=0.4\textwidth]{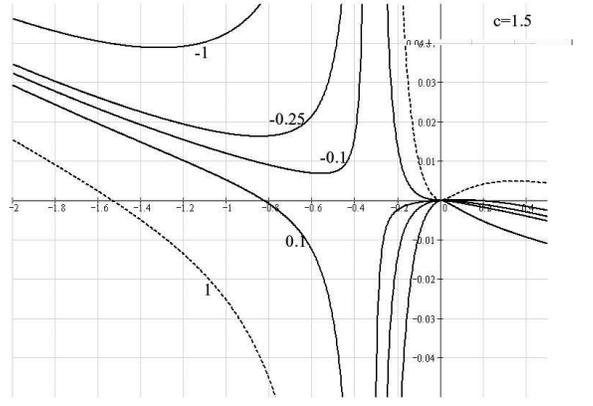}
\caption{The same as in Fig.\,1 but with $c=1.5$. The minima of
$V(\phi)$ are situated in the range $\phi < -1/3$, where $F' < 0$.}
\end{figure}

    We see that the behavior of the system is drastically different in
    different ranges of $\phi$ and depends on the numerical values of
    the initial parameters. We also confirm that gravity alone can stabilize
    the size of extra dimensions, without need for introducing other fields
    with specific forms of potentials. Positive values of the effective
    cosmological constant $\Lambdaef$ (vacuum energy density) were found in
    the range where $F' < 0$. This is a feature of quadratic gravity only:
    as will be seen in the next section, e.g., for cubic gravity one may
    have $\Lambdaef > 0$ where $F' > 0$.

\section{Other theories \label{S4}}

\subsection{Cubic gravity}

    The cubic theory of gravity with
\beq
        F(R) = R + cR^2 + CR^3
\eeq
    is another example of curvature-nonlinear gravity for which
    \eqs (\ref{Kgen}) and (\ref{Vgen}) may be applied. The shape of the
    potential appears to be very sensitive to the values of the parameters
    $c$ and $C$. This property may be used to adjust the potentials in such
    a way that its minimum in the Einstein frame will lead to an extremely
    small but positive value of $\Lambdaef$ in the Jordan frame, and it
    turns out that, unlike quadratic gravity, this can be achieved in the
    range where $F' > 0$. Examples of such a behavior are shown in
    Fig.\,3. Thus cubic gravity can in principle lead to an appropriate
    description of our Universe with compact and stable extra dimensions.
    At large times, when the influence of matter may be neglected compared
    to the effective cosmological constant, we shall obtain de Sitter
    expansion of the three observable dimensions.

\begin{figure}\label{MultiCubea}
    \includegraphics[scale=0.4]{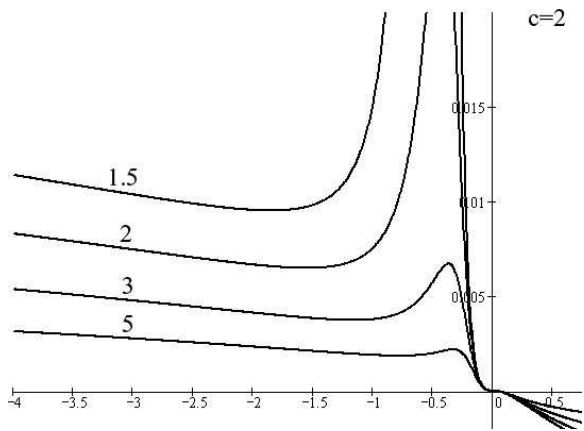}
\caption
   {The potential $V_{\rm Ein}$ for cubic gravity with $d_1=3$ and $c=2$.
    The curves are labelled by values of the parameter $C$.
    Note that $F'(\phi )>0$ near the minima.}
\end{figure}

    The range of positive curvature, $\phi >0$, appears to be of lesser
    interest. It also contains minima but only with $V_{\min} < 0$, see
    Fig.\,4. The states in such minima are metastable (since the minima are
    only local), so that such a space would have a finite lifetime. The
    latter, however, could be made arbitrarily long by choosing the
    parameters.

\begin{figure}\label{MultiCubec-1a}
    \includegraphics[scale=0.4]{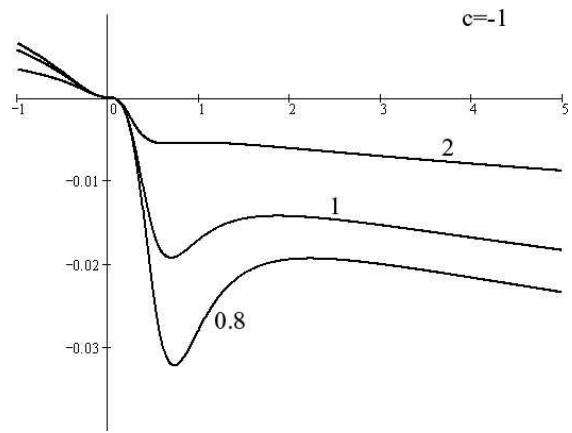}
\caption
   {$V_{\rm Ein}(\phi)$ for cubic gravity with $d_1=3$, $c=-1$.
    Lines are labelled by values of the parameter $C$. Again $F'(\phi )>0$
    near the minima.}
\end{figure}

\subsection{Extensions}

    The method discussed above allows considering wider classes of
    Lagrangians. Let us briefly demonstrate this by adding terms
    proportional to the Ricci tensor squared $R_{AB}R^{AB}$ and the
    Kretschmann scalar $\cK = R_{ABCD}R^{ABCD}$. By common views, these and
    other high-order curvature terms appear due to quantum corrections, and
    it seems natural to include them on equal footing with  $cR^2$. Now the
    action has the form
\bearr
    S = \frac{\cV[d_1]} {2\kappa^2 }\int d^4 x\,\sqrt{^{4}g}
        \e^{d_1 \beta} \bigl(R + cR^2
\nnn \cm
          + c_1 R_{AB} R^{AB} + c_2 {\cK} - 2\Lambda\bigr),  \label{actt}
\ear
    where the internal variables have been integrated out in full analogy
    with \eq (\ref{act1}). For the metric (\ref{ds}), it is easy to obtain
    expressions for $R_{AB}R^{AB}$ and $\cK$:
\bearr
    R_{AB}R^{AB} = \oR\mn\oR\MN + 2d_1 \oR \mn B\MN
        + \e^{-4\beta}\oR_{ab}\oR^{ab},
\nnn\
    + 2\e^{-2\beta} \oR[h] [\DAL\beta + d_1 (\d{\beta})^2 ]
        + d_1 [\DAL\beta + d_1 (\d{\beta})^2 ]^2               \label{Ric2}
\nnn
\\ \lal
    \cK = \overline{\cK}[g] + 4 d_1 B\mn B\MN + \e^{-4\beta}\overline{\cK}[h]
\nnn \qquad
      + 4 \e^{-2\beta} \oR [h] (\d{\beta})^2
         +2 d_1 (d_1-1) [(\d\beta)^2 ]^2 .                     \label{Kre}
\ear

    In the slow-change approximation, in the same manner as in
    \sect\ref{S2}, we obtain the 4D effective Lagrangian
\bearr
    \sqrt{^{4}g} L = \sqrt{^{4}g}\e^{d_1 \beta}
                \Big[R_4 (1 + 2c\phi) +\phi + c_{\rm tot}\phi^2 -2\Lambda
\nnn \qquad
     + (1+2c\phi)f_1 + 2 c_1\phi \DAL\beta
            + 2(c_1 d_1 + 2c_2) (\d\beta)^2 \Big]  \label{L_4a}
\nnn
\ear
    with
\[
        c_{\rm tot} = c + \frac{c_1}{d_1} + \frac{2c_2}{d_1(d_1-1)}.
\]
    The conformal mapping (\ref{trans-g}) leads, after some calculations, to
    the Einstein-frame Lagrangian (\ref{Lgen}) with the kinetic and
    potential terms
\bear                               \label{K_4a}
    K_{\rm Ein}(\phi)
        \eql K^{(2)}_{\rm Ein}(\phi)
        + \frac {c_1 + c_2} {2\phi (1 + 2c\phi)},
\nnn
\\
    V_{\rm Ein}(\phi) \eql -\sign(1 + 2c\phi)
\nnn \nqq \times
     \left[\frac{|\phi |}{d_1 (d_1 -1)}\right]^{d_1/2}
            \frac{c_{\rm tot} \phi^{2}+\phi -2\Lambda}{(1+2c\phi )^2},
\ear
    where the term $K^{(2)}_{\rm Ein}(\phi)$ is taken from (\ref{K}).

    The presence of the parameters $c_1$ and $c_2$ adds freedom in choosing
    the shape of the potential. The kinetic term also acquires a more
    complex form which could significantly affect the field dynamics.  Thus,
    as shown in Ref.\,\cite{Ru04}, zeros of the kinetic term can represent
    stationary values of $\phi$. It means that a field could be captured in
    the vicinity of such points in addition to minima of the potential.  An
    analysis of kinetic terms like (\ref{K_4a}) could lead to possibilities
    of interest, and we hope to return to this point in our future work.

\section{Concluding remarks \label{S5}}

    We have shown that multidimensional gravity alone, even without any
    fields of non-geometric origin and with a very simple choice of the
    geometry, can be a basis for rather complicated phenomena. The assumption
    on nonlinearity of gravity, used in this study, inevitably follows
    from quantum corrections to the Einstein gravity and must be taken into
    account for completeness.

    We have introduced the slow-change approximation in nonlinear
    multidimensional gravity, making the analysis easier. This approximation
    proves to be valid for a wide variety of phenomena where the
    curvature and energy scales are far from Planckian (though Planckian in
    a multidimensional sense, which may mean scales different from those
    known in four-dimensional physics).

    Using some simple examples of nonlinear gravity with an arbitrary
    number of extra dimensions, we have obtained an effective scalar
    field with quite a complex form of the potential. The kinetic term is
    also nontrivial and adds complexity to the effective field dynamics.
    The potentials possess minima, both stable and metastable, thus
    stabilizing the size of extra dimensions. Some of them seem quite
    promising for the description of the present state of the Universe with
    a small positive effective cosmological constant and stable and small
    enough extra dimensions.

    A problem that could not be solved in this paper is that of choosing the
    observational conformal frame, which is a necessary step in confronting
    the theory to observations. For our numerical estimates we identified
    the initial (fundamental) and observational frames. One should, however,
    bear in mind that it is not the only possible choice, and the right one
    should follow from a full underlying theory.

    The precise number of extra dimensions $d_1$ was not very important in
    our study, and we used the only value $d_1 =3$ in numerical
    calculations, though the method is valid for arbitrary $d_1$.
    This number may prove to be more essential in more complex geometries
    and nonlinear gravity theories.

    One of possible applications of this method is the brane world scenario,
    where, in particular, it opens a direct way of obtaining a stabilizing
    radion potential instead of postulating it.

\subsection*{Acknowledgment}

We thank Julio Fabris for a helpful discussion.
KB acknowledges partial financial support from ISTC Project
No. 1655 and DFG Project 436RUS113/807/0-1(R).

\end{document}